\documentclass[prd,showpacs,nofootinbib,preprint, 10 pt]{revtex4}
\usepackage{amsmath,soul,graphicx,xcolor,epsfig,slashed}
\usepackage{appendix}
\usepackage{tabularx}
\usepackage{float}
\usepackage{pdfpages}
\usepackage{epstopdf}
\usepackage{subfigure}
\usepackage{caption}
\usepackage{subcaption}
\usepackage{multirow}
\usepackage{hyperref}
\usepackage{ulem}
\usepackage{breqn}
 
\usepackage{tikz}
\usepackage{tikz-feynman}
\tikzfeynmanset{compat=1.1.0}
\begin{document}
\begin{center}
{\large Symmetry-Breaking Effects on Form Factors and Observables in $B \to K_0^*(1430)\mu^+\mu^-$ Decay
}\\[5mm]

\setlength {\baselineskip}{0.2in}
{Saba Ayub$^{1}$, Saba Shafaq$^{1,2}$, Arslan Sikandar$^3$, M. Jamil Aslam$^3$}\\[5mm]
$^1$~{\it Department of Physics, International Islamic University, Islamabad 44000, Pakistan.}\\
$^2$~{\it National Center for Physics, Islamabad 44000, Pakistan.}\\
$^3$~{\it Department of Physics, Quaid-i-Azam University, Islamabad 45320, Pakistan.}
\\[5mm]
\end{center}
{\bf Abstract}\\[5mm]
In the heavy-quark and large-energy limits, symmetry relations reduce the number of independent form factors governing heavy-to-light $B$-meson decays. Exploiting these relations, the form factors can be parametrized while systematically incorporating symmetry-breaking corrections from perturbative QCD. Using vertex renormalization together with light-cone distribution amplitudes, we compute the vertex and hard-spectator contributions for the $B \to K_0^*(1430)$ transition. We then analyze the impact of these form factors on physical observables, including the branching ratio and lepton polarization asymmetries $(P_L, P_N)$, in $B \to K_0^*(1430)\mu^+\mu^-$. Our results indicate that perturbative corrections induce modest shifts of $\sim 3\%$ in both the branching ratio and the normal lepton polarization asymmetry. Consequently, any significant deviation observed experimentally from these predictions would provide a clear signal of potential New Physics effects.\\


\maketitle

\section{Introduction}

The Standard Model (SM) has achieved remarkable success in describing the fundamental particles and their interactions both theoretically and experimentally. Nevertheless, the search for physics beyond the SM, also known as the new physics (NP) remains one of the primary objectives of contemporary particle physics. While high-energy experiments aim to reveal new particles through direct production, precision studies provide an alternative and complementary approach for uncovering possible deviations from SM predictions. In this context, heavy-flavor physics offers a powerful laboratory for testing the SM with high accuracy, owing to the large amount of available experimental data. A central aspect of these studies is the examination of the unitarity of the Cabibbo-Kobayashi-Maskawa (CKM) matrix, where the $B$-meson decays are always at the forefront.

Among the various $B$-meson decay channels, particularly sensitive probes of the SM arise from flavor-changing neutral-current (FCNC) transitions of the type $b \to s(d)\ell^{+}\ell^{-}$. In the Standard Model, these processes occur only at the loop level and are further suppressed by the CKM matrix elements. Consequently, the corresponding exclusive decay modes, such as 
$B^{\pm}\to K^{(*)\pm}\ell^{+}\ell^{-}$,
$B^{0}\to K^{0}\ell^{+}\ell^{-}$, and
$B_s^{0}\!\to \phi\,\ell^{+}\ell^{-}$,
with $\ell=e,\mu$, have been extensively investigated experimentally \cite{LHCb:2014cxe,LHCb:2014auh,LHCb:2015wdu,LHCb:2015svh,LHCb:2016ykl,LHCb:2021zwz,LHCb:2021xxq}. 
Of particular interest are tests of lepton-flavor universality (LFU) in $B\to K^{(*)}\ell^{+}\ell^{-}$ decays \cite{LHCb:2022vje,LHCb:2022qnv,Smith:2024xgo,CMS:2024syx}. In these observables, the dependence on CKM matrix elements as well as the uncertainties associated with hadronic form factors largely cancel, making them especially clean probes of possible new physics effects. As a result, these processes have been widely studied in a variety of new-physics scenarios; see, for example, Refs. \cite{Celis:2017doq,Buttazzo:2017ixm,Aebischer:2019mlg,Alasfar:2020mne,Isidori:2021tzd,Ciuchini:2022wbq}. Recent experimental measurements indicate that these observables are consistent with the SM predictions within approximately $0.2\sigma$ \cite{Hiller:2003js,Bordone:2016gaq,Mishra:2020orb,Isidori:2020acz}. Motivated by these developments, it is worthwhile to explore other complementary exclusive decays induced by the same FCNC transition $b \to s\ell^{+}\ell^{-}$. In this context, the rare decays $B \to S\,\ell^{+}\ell^{-}$, where $S=(f_0,a_0,K_0^*)$ denotes a scalar meson, provide an additional probe of the underlying flavor dynamics.
 
The study of light scalar mesons with masses below $1.5$ GeV serves as an intriguing subject of study due to their non trivial internal structure. While they are usually viewed as conventional quarks-antiquark states \cite{Chen:2021oul}, alternate descriptions have been proposed depending upon their masses and observed properties. These include tetra-quark configurations \cite{Cheng:2019tgh}, meson-meson molecular states \cite{Han:2013zg}, and less convincingly, glueball states \cite{Bharucha:2015bzk}. Although some of these models are quite successful in explaining certain experimental features, none of them provides a full experimentally observed consistent description. As a consequence, the internal structure of light scalar mesons remains an open question and continues to attract considerable interest in hadron physics. The meson $K_0^*(1430)$, which constitutes the main focus of the present study, is commonly interpreted as a predominantly $s\bar{q}$ or $q\bar{s}$ state in many phenomenological analyses. However, its classification is still subject to debate, with two commonly discussed scenarios in the literature. In the first scenario, $K_0^*(1430)$ is treated as an excited state associated with a lighter scalar ground state below $1\,\mathrm{GeV}$. In the second scenario, it is regarded as the lowest-lying scalar state, while the light scalar nonet below $1\,\mathrm{GeV}$ is interpreted as a set of tetraquark bound states. A detailed discussion of these possibilities can be found in Refs.~\cite{Du:2004ki,Lu:2006fr}.
To probe the quark-antiquark structure of $K_0^*(1430)$, the semileptonic weak decay $B \to K_0^*(1430)\ell^+\ell^-$ provides a relatively clean channel compared to purely hadronic decay modes, reducing uncertainties associated with strong interactions.  As a weak decay, the key inputs to the SM calculation are the hadronic matrix elements of the weak currents, parameterised by form factors. The form factors are functions of the four-momentum transfer, $q$, between $B$ and $K_0^*\left(1430\right)$ and depend on the strong interaction effects that bind the quarks inside the mesons, and hence clarify the internal structure of the $K_0^*\left(1430\right)$ meson. Several theoretical approaches are used in literature including simple quark model \cite{Wirbel:1985ji}, light front approach \cite{Cheung:1995ub, Zhang:1994hg, Choi:1999nu}, QCD sum rules \cite{Shifman:1978by,  Novikov:1981xi}, light cone sum rules \cite{Balitsky:1989ry, Braun:1988qv, Chernyak:1990ag} and perturbative QCD factorization approach \cite{Keum:2000ph, Keum:2000wi, Lu:2000em} for the precise measurements of these form factors.  

As the form factors are non-perturbative, model-dependent quantities that dominate theoretical uncertainties in $B$-meson decay predictions \cite{Gubernari:2018wyi,Beneke:2000wa,Beneke:2001at,Hatanaka:2008xj,Paracha:2007yx,Momeni:2017vae}, particularly in the low momentum-transfer region. Effective field theories (EFT) allow certain symmetries to reduce the number of independent form factors. In particular, heavy-quark symmetry (HQS), applicable for mesons containing a heavy quark, provides symmetry relations \cite{Isgur:1989vq,Isgur:1990yhj,Grinstein:1990mj,Neubert:1993mb,Charles:1998dr,Grozin:1996pq,Georgi:1990um} that are not explicit in full QCD. These relations allow form factors to be expressed in terms of a reduced set of universal Isgur–Wise functions, minimizing the number of hadronic parameters.  

Although the form-factor structures for $B\to K^*$ \cite{Khodjamirian:2012rm,Khodjamirian:2010vf} and 
 $B\to D^*$ \cite{Grozin:1996pq,Gubernari:2018wyi,Georgi:1990um,Shafaq:2019kte} have been studied extensively; further efforts are continued to achieve higher precision, particularly in the kinematical situations where the outgoing degree of freedom carries a large amount of energy $(E)$. In such large recoil regimes relevant to semi-leptonic decays $B$ to $(\pi,\rho, K^*)\ell^+\ell^-$, the large-energy-effective-theory (LEET) plays an important role by providing an essential theoretical framework. Within this approach, form factors can be factorized using HQS, for the initial state heavy meson and LEET for energetic final state light meson, into hard and soft parts \cite{Beneke:2000ry,Beneke:2003xr,Beneke:2003pa}. The soft contributions correspond to gluon interactions of order $\Lambda_Q{}_C{}_D/m_b$ while the hard spectator part, involving the spectator quark, is of order $m_b\Lambda_Q{}_C{}_D$. In the decay $B\to V$, LEET reduces the seven form factors into two in the large recoil limit. Furthermore, the large-energy of the final state meson further highlights the importance of perturbative corrections. To compute these corrections, one needs a suitable factorization scheme to separate the perturbative and the non-perturbative parts. One such factorization scheme is introduced in Eq.(\ref{fact}). In this framework, the hard gluon vertex corrections are absorbed in the coefficients $C_i$ of the soft-form factors. Additionally, at order $1/m_b$, all end-point singularities \cite{Buchalla:1995vs} that appear in the hard-spectator interactions are also absorbed in the soft-form factors as they respect heavy quark symmetry. On the other hand, the corrections that violate these symmetries are treated separately and explicitly incorporated in the form factors. 
 
The main objective of this study is the calculation of hard-spectator corrections together with vertex renormalization for the decay $B \to K_0^*(1430)\,\ell^+\ell^-$. In the large-energy limit, heavy-quark symmetry reduces the three form factors defined through the relevant matrix elements to a single universal function, $\xi_{K_0^*}(E_F)$. Symmetry-breaking corrections to this form factor are evaluated explicitly using vertex renormalization and hard-spectator interactions. An accurate determination of these corrections is essential not only for reliable theoretical predictions but also for guiding future high-precision measurements at experiments such as LHCb and Belle II, where rare semileptonic $B$-meson decays provide sensitive probes of hadronic dynamics and potential new-physics effects. After quantifying these corrections, we analyze their impact on physical observables, including the branching ratio and various lepton-polarization asymmetries in these decays.

This work is organized as follows: In the Sec. \ref{TFW}, we have discussed the theoretical framework used to evaluate the form factors under the symmetry relation. Sec. \ref{VH-corrections} is divided into two parts: the first part discusses the correction to the vertex, while the second portion is dedicated to hard spectator correction at order $\alpha_s$ using the light cone distribution amplitudes (LCDAs). These form factors serve as important inputs for the analysis of the branching fraction and lepton polarization asymmetry. Our numerical and analytical results and their comparison with the theoretical predictions are presented in Sec. \ref{Tanalysis}. Finally, Sec. \ref{conclusion} is dedicated to the conclusion. 

\section{Theoretical Framework}\label{TFW}
\subsection{Weak Effective Hamiltonian}
The weak effective Hamiltonian for the rare $B$ meson decays can be obtained by integrating out the heavy degrees of freedom, such as the $W$-boson, $t$-quark and the Higgs boson \cite{Bauer:2002aj}. This approach is known as the operator product expansion (OPE), where the short distance (SD) effects are rendered in the Wilson coefficients $\mathcal{C}_i$, leaving the operators $\mathcal{O}_i$ describing the physics at a long distance (LD). Implementing this, the weak effective Hamiltonian can be written as :

\begin{eqnarray}
 H_{eff}=-\frac{4 G_{F}}{\sqrt{2}}\lambda_{t}\left[\sum_{i=1}^{6}C_{i}(\mu)O_{i}(\mu)+\sum_{i=7,9,10}C_{i}(\mu)O_{i}(\mu) \right].\label{H11}
\end{eqnarray}
In Eq. (\ref{H11}) $\lambda_{t}=V_{tb}V_{ts}^{\ast}$ are the CKM matrix elements, $G_{F}$ is the Fermi coupling constant, $C_{i}$ are 
the Wilson coefficients, and $O_{i}$ are the SM with operators with $V-A$ structure. For $B\to K_0^\ast\left(1430\right) \ell^{+}\ell^{-}$ decays in the SM, the operators $O_{7,\; 9,\; 10}$ and their corresponding WCs $C_{7,\;, 9\; 10}$ will contribute. These operators have the form 
\begin{eqnarray}
 O_{7} &=&\frac{e}{16\pi ^{2}}m_{b}\left( \bar{s}\sigma _{\mu \nu }P_{R}b\right) F^{\mu \nu }\,,  \notag \\
O_{9} &=&\frac{e^{2}}{16\pi ^{2}}(\bar{s}\gamma _{\mu }P_{L}b)(\bar{\ell}\gamma^{\mu }\ell)\,,  \label{op-form} \\
O_{10} &=&\frac{e^{2}}{16\pi ^{2}}(\bar{s}\gamma _{\mu }P_{L}b)(\bar{\ell} \gamma ^{\mu }\gamma _{5} \ell)\,.  \notag
\end{eqnarray}
Specifically, the operator $O_{7}$ describe the interaction of $b$ and $s$ quarks with the emission of a photon, whereas $O_{9,\; 10}$ correspond to the interaction of these quarks with charged leptons through (almost) same Yukawa couplings.
 
The WCs given in Eq.(\ref{H11}) encode the short distance (high momentum) contributions, and these are calculated using the perturbative approach. The contributions from current-current,  QCD penguins and chromomagnetic operators $O_{1-6,8}$, \textit{i.e.,} 
\begin{eqnarray}
O_{1}&=&\left(\bar{s}_ic_j\right)_{V-A}\left(\bar{c}_j b_i\right)_{V-A},\;\;\quad\quad\quad\quad\quad O_{2}=\left(\bar{s}_ic_i\right)_{V-A}\left(\bar{c}_j b_j\right)_{V-A},\notag\\
O_{3}&=&\left(\bar{s}_ib_i\right)_{V-A}\sum_q\left(\bar{q}_jq_j\right)_{V-A},\quad\quad\quad\quad O_{4}=\left(\bar{s}_ib_j\right)_{V-A}\sum_q\left(\bar{q}_jq_i\right)_{V-A},\notag\\
O_{5}&=&\left(\bar{s}_ib_i\right)_{V-A}\sum_q\left(\bar{q}_jq_j\right)_{V+A},\quad\quad\quad\quad O_{6}=\left(\bar{s}_ib_j\right)_{V-A}\sum_q\left(\bar{q}_jq_i\right)_{V+A},\notag\\
O_{8}&=&\frac{g_s m_b}{8\pi^2}\bar{s}_i\sigma^{\mu\nu}\left(1+\gamma_5\right)T^a_{ij}b_jG^a_{\mu\nu},\label{O1to8}
\end{eqnarray}
have been unified in the WCs $C_{9}^{\text{eff}}$ and $C_{7}^{\text{eff}}$, and their explicit expressions are given as follows \cite{Beneke:2001at, Greub:2008cy}:
\begin{eqnarray}
 C_{7}^{\text{eff}}(q^{2})=C_{7}-\frac{1}{3}\left(C_{3}+\frac{4}{3}C_{4}+20C_{5}+\frac{80}{3}C_{6}\right)-\frac{\alpha_{s}}{4\pi}\bigg[\left(C_{1}-6C_{2})F^{(7)}_{1,c}(q^{2})+C_{8}F^{7}_{8}(q^{2}\right)\bigg]\notag\\
 C_{9}^{\text{eff}}(q^{2})=C_{9}+\frac{4}{3}\left(C_{3}+\frac{16}{3}C_{5}+\frac{16}{9}C_{6}\right)-h(0,q^{2})\left(\frac{1}{2}C_{3}+\frac{2}{3}C_{4}+8C_{5}+\frac{32}{3}C_{6}\right)\notag\\
-\left(\frac{7}{2}C_{3}+\frac{2}{3}C_{4}+38C_{5}+\frac{32}{3}C_{6}\right)h(m_{b},q^{2})+\left(\frac{4}{3}C_{1}+C_{2}+6C_{3}+60C_{5})h(m_{c},q^{2}\right)\notag\\
-\frac{\alpha_{s}}{4\pi}\bigg[C_{1}F^{(9)}_{1,c}(q^{2})+C_{2}F^{(9)}_{2,c}(q^{2})+C_{8}F^{(9)}_{8}(q^{2})\bigg]\label{WC3}
\end{eqnarray}
The WC given in Eq. (\ref{WC3}) involves the functions $h(m_{q},s)$ with $q=c,b$, functions $F^{7,9}_{8}(q^{2})$, and $F^{(7,9}_{1,c}(q^{2})$ are
defined in \cite{Beneke:2001at, Greub:2008cy,Bharucha:2015bzk}.

The numerical values of the Wilson coefficients $C_{i}$ for $i=1,...,10$ at $\mu\sim m_{b}$ scale are presented in Table \ref{wc table}.
\begin{table*}[tbh]
\caption{The Wilson coefficients $C_{i}$ at the scale $\mu\sim m_{b}$ in the SM.}
\begin{tabular}{cccccccccc}
\hline\hline
$C_{1}$&$C_{2}$&$C_{3}$&$C_{4}$&$C_{5}$&$C_{6}$&$C_{7}$&$C_{9}$&$C_{10}$
\\ \hline
  \  -0.263 \  &  \  1.011  \ & \ 0.005 \ &   \ -0.0806   \ &   \ 0.0004  \ &   \ 0.0009   \ &   \ -0.2923  \ &   \ 4.0749  \ &  \ -4.3085 \ \\
\hline\hline
\end{tabular}
\label{wc table}
\end{table*}

\subsection{Matrix elements and form factors}
Sandwiching the effective Hamiltonian between initial state $B$ and final state $K_0^\ast\left(1430\right)$ involves the following matrix elements: 
\begin{equation}
\langle {K^*_0}(p_F)|\bar{q}\gamma^\mu\gamma_5b|B(p_B)\rangle\;,\quad\quad \langle {K^*_0}(p_F)|\bar{q}\sigma^\mu{}^\nu \gamma_5q_\nu b|B(p_B)\rangle\;,
\end{equation}
which in terms of form factors $f_\pm$ and $f_T$ can be expressed as:
\begin{eqnarray}
\langle {K^*_0}(p_F)|\bar{q}\gamma^\mu\gamma_5b|B(p_B)\rangle&=&f_+(q^2)(p^\mu_B+p^\mu_F)+f_-(q^2)(p^\mu_B-p^\mu_F)\;,   
\label{D}\\
\langle {K^*_0}(p_F)|\bar{q}\sigma^\mu{}^\nu \gamma_5q_\nu b|B(p_B)\rangle&=&\frac{i f_T(q^2)}{m_B+m_{K^*_0}}\left[(p^\mu_B+p^\mu_F)q^2-(m_B^2-m^2_{K^*_0})q^\mu\right]\;.
\label{E}
\end{eqnarray}
Here, $m_B\left(m_{K_0^*}\right)$ and $p_B\left(p_F\right)$ is the mass and momentum of $B\;\left(K^*_0\left(1430\right)\right)$ meson, respectively, and $q^2=(p_B-p_F)^2$ is the momentum transfer. In terms of the masses and $q^2$, the energy of the final state particle is defined as:
\begin{equation}
E_F=\frac{m^2_B-m^2_{K^*_0}-q^2}{2m_B}\;.
\end{equation}
Since we are interested in the large-recoil region, where the momentum transfer is small, i.e., $q^2 \ll m_B^2$ and $m_{K_0^*}^2 \ll m_B^2$, the final-state meson carries a large energy. In general, if the mass of the final-state particle is below $1\,\mathrm{GeV}$, its energy $E_F$ is typically of order $\sim m_B/2$, making $E_F$ a natural expansion parameter in the large-recoil limit. 
In the present case, however, the scalar meson $K_0^*(1430)$ has a mass of about $1.43\,\mathrm{GeV}$, which lies well above the hadronic scale $\Lambda_{\rm QCD}$. Therefore, special care is required when separating perturbative and non-perturbative contributions in the form-factor analysis. In particular, it is important to retain terms of order $m_{K_0^*}^2/m_B^2$ in the expansion.

In terms of the velocity $v$ of the heavy quark, the momentum of  $B$-meson is given as $p_B^\mu=m_Bv^\mu$. $B$-meson is a bound state of heavy and light quark, therefore, in terms of $v$ the momentum of heavy quark will read as: 
\begin{equation}
p^\mu_Q=m_Q v^\mu+k^\mu\;,    
\end{equation}
where $k$, representing the off-shellness of the heavy quark, denotes the residual momentum that scales as $k^{\mu} \sim \Lambda_{\rm QCD} \ll m_Q$. In the rest frame of the parent meson, the four-velocity is defined as $v^{\mu}=(1,0,0,0)$. 

In the large-energy effective theory (LEET), the energy of the final-state particle is used as an expansion parameter. Therefore, in the large-recoil regime it is convenient to employ light-cone coordinates. These can be introduced using the lightlike four-vectors $n_+^{\mu}=(1,0,0,1)$ and $n_-^{\mu}=(1,0,0,-1)$, which satisfy the conditions $n_+^2=n_-^2=0$ and $n_+\!\cdot n_-=2$. 

The momentum of the final-state light meson can then be expressed in terms of these light-cone vectors as
\begin{equation}
p_F^{\mu}=E\,n_+^{\mu}+\frac{m_{K_0^*}^2}{4E}\,n_-^{\mu},
\end{equation}
where, $p_F^2=m_{K_0^*}^2$ and $E$ denotes the energy of the final-state meson in the large-recoil limit \cite{Ebert:2001pc,Sikandar:2019qyb,Sikandar:2022iqc}. The corresponding relations for the three-momentum and on-shell energy of the final-state scalar meson are given by
\begin{equation}
E_F = E\left(1+\frac{m_{K_0^*}^2}{4E^2}\right), \qquad 
|\vec{\Delta}|\equiv \Delta =  \sqrt{E_F^2-m_{K_0^*}^2}
      = E\left(1-\frac{m_{K_0^*}^2}{4E^2}\right).
\end{equation}
Accordingly, the momentum of the light $\left(s\right)$ quark in the final state can be written as
\begin{equation}
p_s^{\mu}=E\,n_+^{\mu}+\frac{m_{K_0^*}^2}{4E}\,n_-^{\mu}+k^{\prime\mu} =\Delta\,n_+^{\mu}+\frac{m^2_{K_0^*}}{2E}\,v^{\mu}+k^{\prime \mu},
\end{equation}
where $k^{\prime \mu}$ denotes the residual momentum of order $\sim\Lambda_{\rm QCD}\ll E$. 
It is then convenient to write
\begin{equation}
(p_B+p_F)^\mu
= m_B\left(1+\frac{m_{K_0^*}^2}{2Em_B}\right)v^\mu
  + \Delta\,n_+^\mu ,
\qquad
q^\mu \equiv (p_B-p_F)^\mu
= m_B\left(1-\frac{m_{K_0^*}^2}{2Em_B}\right)v^\mu
  - \Delta\,n_+^\mu .
\end{equation}

The hadronic matrix elements describing the $B\to K_0^*$ transition in the large-recoil limit can be evaluated using the standard framework of heavy-quark effective theory (HQET). In this approach, the matrix elements can be written in terms of universal functions through the relation \cite{Neubert:1993mb}:
\begin{equation}
\langle {K^*_0}(p_F)|\bar{q}\Gamma b|B(p_B)\rangle=\mathrm{Tr}\left[A_S(E_F)\mathcal{\overline{M}}_{K^*_0}\Gamma \mathcal{M}_B\right]\;,
\label{AA}
\end{equation}
where $\Gamma$ denotes an arbitrary Dirac structure. The matrices $\mathcal{\overline{M}}_{K^*_0}$ and $\mathcal{M_B}$ are spin projectors for $K_0^*$ and $B$ mesons, respectively. In terms of $v$ and $n$, they are of the form:
\begin{equation}
\mathcal{\overline{M}}_{K^*_0}=\dfrac{\slashed{v}\slashed{n_+}}{2},\hspace{0.5cm}\mathcal{M}_B=-\dfrac{1+\slashed{v}}{2}\gamma_5\;.
\label{B}
\end{equation}
The function $A_S(E_F)$ contains the LD QCD dynamics and is independent of the Dirac structure. For the $K^*_0$ meson in the final state, it becomes
\begin{equation}
A_S(E_F)=2E_F\xi_{K^*_0}(E_F)\;,
\label{C}
\end{equation}
where $\xi_{K_0^*}(E_F)$ denotes the universal soft form factor. 
The symmetry relations among the form factors can be obtained by substituting 
Eqs.~(\ref{B}) and (\ref{C}) into Eq.~(\ref{AA}) and evaluating the traces 
for the relevant Dirac structures, namely 
$\Gamma=\gamma^\mu\gamma_5$ and $\Gamma=\sigma^{\mu\nu}\gamma_5q_\nu$. This gives
\begin{eqnarray}
 \langle {K^*_0}(p_F)|\bar{q}\gamma^\mu\gamma_5b|B(p_B)\rangle&=&2E_F\xi_{K^*_0}(E_F)n_+^\mu\;, \nonumber\\
\langle {K^*_0}(p_F)|\bar{q}\sigma^\mu{}^\nu \gamma_5q_\nu b|B(p_B)\rangle&=&2E_F\xi_{K^*_0}(E_F)\left[(m_B-E_F)n_+^\mu-m_B\left(1-\frac{m^2_{K^*_0}}{2Em_B}\right)v^\mu\right]\;,
\label{G}
\end{eqnarray}
demonstrating that in the large-recoil limit, the matrix elements of $B\to K^*_0\left(1430\right)$ can be parameterized in terms of a single invariant function $\xi_{K_0^*}(E_F)$. Consequently, by comparing Eq. (\ref{D}) and Eq. (\ref{E}) with Eq. (\ref{G}), one obtains
\begin{equation}
f_+(q^2)=\left(1-\frac{m^2_{K^*_0}}{2Em_B}\right)\frac{E_F}{\Delta}\xi_{K^*_0}(E_F)\;.    
\end{equation}
This expression can be further simplified by expanding in the heavy-quark limit in powers of $m_{K_0^*}/E$. It should be noted that, due to the relatively large mass of the $K_0^*$ meson, the terms of order $m_{K_0^*}^2/m_B^2$ cannot be neglected and are therefore retained in the expansion. However, upon neglecting the power correction term $m_{K_0^*}^2/m_B^2$, one finds that the results reported in \cite{Beneke:2000wa} are recovered. The resulting expression for the form factor $f_+(q^2)$ is given by
\begin{equation}
f_+(q^2)=\left(1+\frac{m_{K_0^*}^2}{m_B^2}\right)\xi_{K_0^*}(E_F).
\end{equation}

Similarly, the remaining form factors $f_-(q^2)$ and $f_T(q^2)$ can be written as
\begin{eqnarray}
f_-(q^2)&=&-\left(1+\frac{3m_{K_0^*}^2}{m_B^2}\right)\xi_{K_0^*}(E_F),\\
f_T(q^2)&=&\frac{m_B+m_{K_0^*}}{m_B}\left(1+\frac{2m_{K_0^*}^2}{m_B^2}\right)\xi_{K_0^*}(E_F).
\end{eqnarray}
The above expressions, derived using HQS, allow all form factors to be expressed in terms of the universal soft form factor $\xi_{K_0^*}(E_F)$. Consequently, one obtains the relation
\begin{equation}
f_+(q^2)
=-\left(1-\frac{2m_{K_0^*}^2}{m_B^2}\right)f_-(q^2)
=\frac{m_B}{m_B+m_{K_0^*}}\left(1-\frac{m_{K_0^*}^2}{m_B^2}\right)f_T(q^2)
=\left(1+\frac{m_{K_0^*}^2}{m_B^2}\right)\xi_{K_0^*}(E_F).
\label{A}
\end{equation}

These relations are valid for the soft (nonfactorizable) contributions to the form factors in the large-recoil region. They receive perturbative corrections of order $\mathcal{O}(\alpha_s)$ and power-suppressed corrections of order $\mathcal{O}(1/m_b)$, which will be computed in the next section.
\section{Symmetry Breaking Corrections}\label{VH-corrections}
\begin{figure}[h!]
    \centering
    \subfigure[]
{\includegraphics[width=6cm, height=4.65cm]{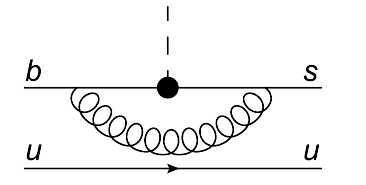}}
\subfigure[]{\includegraphics[width=11cm, height=5cm]{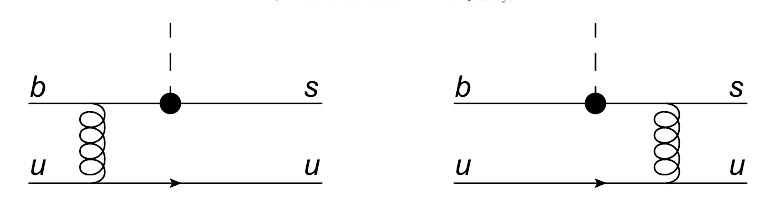}}
    \caption{Vertex and Hard Spectator Corrections in $B$ to $K^*_0(1430)$ decays}
    \label{diagrams}
    \end{figure}

In the previous section, we showed that HQS allows the various form factors to be related to each other [cf. Eq.~(\ref{A})]. However, at $\mathcal{O}(\alpha_s)$ these symmetry relations receive corrections arising from vertex and hard-spectator interactions, as illustrated in Fig.~(\ref{diagrams}). In this section, we evaluate these two types of corrections separately.

Since the hard and soft contributions from Fig.~(\ref{diagrams}-b) cannot be uniquely separated and exhibit logarithmic divergences, a suitable factorization framework is required. For a detailed discussion of this approach, we refer the reader to the works of Beneke \textit{et al.}~\cite{Beneke:2001at} and Refs.~\cite{Chernyak:1990ag,Khodjamirian:1997ub,Bagan:1997bp}.
The factorization formula for the heavy to light transitions at large recoil valid at leading order in $1/m_b$ can be summarized as:
 \begin{equation}
    f_i(q^2)=C_i\xi_{K^*_0}(E_F)+\Phi_B\otimes \mathcal{T}^\Gamma\otimes \Phi_{K^*_0}, 
    \label{fact}
 \end{equation}
 where $\xi_{K^*_0}(E_F)$ is the soft form factor to which the above calculated symmetry relations are applied. The quantity $\mathcal{T}^\Gamma$ represents a hard scattering kernel which is convoluted with light cone distribution amplitudes $\Phi_B$ and $\Phi_{K^*_0}$ of $B$ and $K^*_0$ mesons, respectively. The coefficient $C_i=1+\mathcal{O}(\alpha_s)$ accounts for the vertex and hard-spectator renormalization effects, which we will compute below in terms of the vertex and the hard-spectator interaction diagrams. 
\subsection{Vertex Correction}
The one-loop diagram, Fig. (\ref{diagrams}-a), contains both the ultraviolet (UV) and infrared (IR) divergences. The UV divergences are treated by dimensional regularization $(d=4-2\epsilon)$, whereas the infrared divergences are handled by introducing a small parameter $\lambda$, \textit{i.e.}, the mass term for gluon. Employing the standard Passarino-Veltman reduction techniques and keeping the finite mass term $m^2_{K^*_0}$, the one loop results for Dirac structure $\Gamma$ can read as \cite{Beneke:2000wa, Beneke:2001at}:
\begin{eqnarray}
\bar{u}(p')\Gamma(p',p)u(p)&=&\frac{\alpha_sC_F}{4\pi}\bar{u}(p')\Bigg[\Bigg\{-\frac{1}{2}\text{ln}\left(\frac{\lambda^2m_{b}^{2}}{m_{b}^{2}-q^2}\right)-2\text{ln}\left(\frac{\lambda^2m_{b}^{2}}{(m_{b}^{2}-q^2)^2}\right) \notag\\
&-&2\text{Li}_2\left(\frac{q^2}{m_{b}^2}\right)+\frac{2m^2_B}{m^2_B-q^2}L-3-\frac{\pi^2}{2}\Bigg\}\Gamma
+\frac{1}{4}\left\lbrace\frac{1}{\hat{\epsilon}}+3-\text{ln}\left(\frac{m_b ^2}{\mu^2}\right)-L'\right\rbrace \gamma^\alpha\gamma^\beta\Gamma\gamma_{\beta}\gamma_{\alpha}\notag \\
&+&\frac{1}{2q^2}\left\lbrace 1-\frac{ m_B}{2 E_F}L\right\rbrace \gamma^{\alpha}\slashed{p}\Gamma\slashed{p^{'}}\gamma_{\alpha}+\frac{1}{2q^2}\left\lbrace 1-L^{\prime}\right\rbrace m_b \gamma^{\alpha}\slashed{p}\Gamma \gamma_{\alpha}\notag \\
&-&\left.\frac{1}{2q^2}\left\lbrace 2-L^{\prime\prime }\right\rbrace m_b \Gamma \slashed{p'}\right] u(p)\;,
\end{eqnarray}
with $\bar{u}(p')$ and $u(p)$ denoting the external Dirac spinors for light and heavy quarks, respectively. The above expression is derived by using the on shell relations $\bar{u}(p')\slashed{p'}=0 $ and $ \slashed{p}u(p)=m_bu(p)$. The pole $\dfrac{1}{\hat{\epsilon}}$ is defined as $\dfrac{1}{\epsilon} -\gamma_E +\ln 4 \pi $ 
and is subtracted out in the $\overline{\text{MS}}$ scheme. Here, the momentum transfer $q^2=m_B^2+m_{K^*_0} ^2-2m_B E_F$ as mentioned earlier, and for convenience, abbreviations are introduced for the remaining terms given as:  
\begin{eqnarray*}
L&=&-\frac{2E_F}{m_B-2E_F +\frac{m_{K^*_0} ^2}{m_B}}\ln\left(\frac{2E_F}{ m_B}-\frac{m_{K^*_0} ^2}{ m_B^2}\right),\nonumber\\
L^{\prime}&=&L\left(1-\frac{m_{K^*_0} ^2}{2E_F m_B}\right),\nonumber\\
L^{\prime\prime}&=&L\left(4-\frac{m_B  }{E_F}-\frac{2m_{K^*_0} ^2}{E_F m_B}\right)\;. \nonumber
\label{vertex}
\end{eqnarray*}
It is helpful to fix the renormalization convention for soft form factors by imposing the following condition, which holds exactly to all orders in perturbation theory \cite{Beneke:2000wa}
\begin{equation}
f_+(q^2)=\left(1+\frac{m^2_{K^*_0}}{m_B^2}\right)\xi_{K^*_0}(E_F)
\label{para}
\end{equation}
Once the factorization scheme is specified, for any given Dirac current $\Gamma$, calculate the contributions of $\mathcal{O}(\alpha_s)$ by inserting it in Eq.(\ref{AA}). Applying the renormalization convention in Eq.(\ref{para}) and then comparing the result with the form factor expressions in Eq.(\ref{G}). The results obtained are
\begin{align}
f_{-}(q^2)
&= -\xi_{K_0^*}\Bigg[
\frac{\left(1 + \frac{m_{K_0^*}^2}{m_B^2}\right)}{\left(1 - \frac{2m_{K_0^*}^2}{m_B^2}\right)}
- \frac{\alpha_s C_F E_F}{4\pi}
\Bigg\{
\frac{m_{K_0^*}^4+4 \Delta E m_{K_0^*}^2}{2 E^2 \Delta}
\left(
\dfrac{1-\dfrac{m_B}{2 E_f} L}{2 q^2} -\dfrac{2-L^{\prime\prime}}{2 q^2}
\right) \nonumber \\
&\qquad
- \frac{m_{K_0^*}^2}{m_B E\,\Delta}
\left(
\frac{2m_B^2}{m_B^2 - q^2}L- L^\prime
- \ln\!\frac{m_b^2}{\mu^2} + 3
\right)
+ 4m_B \left(\frac{1 - L^\prime}{q^2}\right)
- \frac{m_{K_0^*}^4}{ E^2\Delta}
\left(
\frac{1 - \frac{m_B }{2E_F}L}{2 q^2}
\right)
\Bigg\}
\Bigg]\;,
\end{align}

and the expression for $f_T$ is given as
\begin{align}
f_T(q^2)=&\xi_{K_0^*}\Bigg[
\left(\frac{m_B+m_{K_0^*}}{m_B}\right)
\left(\frac{m_B^2+m_{K_0^*}^2}{m_B^2-m_{K_0^*}^2}\right)
\nonumber\\[4pt]
&+\frac{\alpha_s C_F E_F}{4\pi}
\frac{ m_B(m_B+m_{K_0^*}) }{(E_F m_B - m_{K_0^*}^2)}\times
\Bigg\{
\left(
\frac{1}{m_B}+\frac{m_{K_0^*}^2}{2E\Delta m_B}
-\frac{m_{K_0^*}^2}{E m_B^2}
-\frac{m_{K_0^*}^4}{4E^2\Delta m_B^2}
\right)\frac{2m_B^2}{m_B^2-q^2}L
\nonumber\\[6pt]
&+\left(
-4\Delta-\frac{2m_{K_0^*}^2}{E}
+\frac{4m_{K_0^*}^2}{m_B}
\right)\frac{1-\frac{m_B}{2E_F}L}{2q^2}
+\left(
4m_B+\frac{m_B m_{K_0^*}^2}{E\Delta}
-\frac{3m_{K_0^*}^2}{E}
-\frac{m_{K_0^*}^4}{2E^2\Delta}
\right)\frac{1-L^\prime}{2q^2}
\nonumber\\[6pt]
&+\left(
-2\Delta+\frac{m_{K_0^*}^4}{2E\Delta m_B}
-\frac{m_{K_0^*}^4}{4E^2\Delta}
+\frac{m_{K_0^*}^2}{m_B}
-\frac{m_{K_0^*}^2}{E}
\right)\frac{2-L^{\prime\prime}}{2q^2}
\Bigg\}\Bigg]
\end{align}
\subsection{Hard-Spectator Interaction}
The form factors calculated above will receive further corrections from the interactions with the spectator quarks denoted as $\Delta f_{-,T}$. These hard-spectator corrections arise at $\mathcal{O}\left(\alpha_s\right)$ when the active quark interact with the spectator quark as drawn in Fig. \ref{diagrams}-b. 
The momentum of the heavy $b-$quark and the spectator quark in the $B$-meson are
\begin{equation}
p^\mu=m_bv^\mu, \hspace{2cm}l^\mu=\frac{l_+}{2}n^\mu_++l^\mu_\perp+\frac{l_-}{2}n^\mu_-
\label{hseq1}
\end{equation}
The momentum of $s-$quark and the spectator quark in the final state $K_0^*$ meson are given by
\begin{equation}
k^\mu_1=uE_Fn^\mu_-+k^\mu_\perp+\left(\frac{\vec{k}^2_\perp}{4uE_F}+\frac{m^2_{K^*_0}}{4uE_F}\right)n^\mu_+\;,
\hspace{1.5cm}
k^\mu_2=\bar{u}E_Fn^\mu_--k^\mu_\perp+\left(\frac{\vec{k}^2_\perp}{4\bar{u}E_F}+\frac{m^2_{K^*_0}}{4uE_F}\right)n^\mu_+\;,
\label{hseq2}
\end{equation}
where $\bar{u}=1-u$. All components of the spectator momenta $l$, $k_1$, and $k_2$ are of order $\Lambda_{\rm QCD}$. The momentum of the final-state meson satisfies $(k_1+k_2)^2\sim m_{K_0^*}^2$, which would otherwise scale as $\Lambda_{\rm QCD}^2$ and could be neglected at leading order. Consequently, the exchange of a hard gluon with momentum of order $m_B\Lambda_{\rm QCD}$ becomes relevant in the $n_-$ direction.

The hard-spectator contribution to the heavy-to-light current matrix element can therefore be expressed in terms of the convolution formula
\begin{equation}
\langle K^*_0|\bar{q}\Gamma b|B\rangle=\frac{4\pi\alpha_s C_F}{N_C}\int^1_0 du\int^\infty_0 dl_+ \mathcal{M}^B_{jk} \mathcal{M}^{k^*_0}_{li} \mathcal{T}^\Gamma _{ijkl}\;.
\label{HS matrix element}
\end{equation}
Here $\Gamma=\gamma^\mu\gamma_5$, $\sigma^\mu{}^\nu \gamma_5q_\nu$ and  $\mathcal{T}^\Gamma _{ijkl}$ denote the hard scattering amplitude computed from the Feynman diagrams shown in Fig. \ref{diagrams}-b. The quantities $\mathcal{M}^B$ and $\mathcal{M}^{K^*_0}$ are the two light-cone projectors that contain the non-perturbative dynamics of initial and final bound state mesons. The $K^*_0$ meson projector is given by \cite{Beneke:2000wa,Hai:2006}
\begin{equation}
\mathcal{M}^{K_0^*}_{li}
=
\frac{\bar f_{K_0^*}}{4}
\Bigg[
\not{p_F}\, \phi_{K_0^*}(u)
+ m_{K_0^*}\, \phi_{K_0^*}^s(u)
+ m_{K_0^*}\,
\sigma_{\mu\nu}\, p_F^\mu n^\nu\,
\frac{\phi_{K_0^*}^\sigma(u)}{6}
\Bigg]_{li}\;,
\label{Scalar project}
\end{equation}
where $f_{K^*_0}$ is $K_0^*$ decay constant, $\phi_{K_0^*}(u)$ is a twist-2, $\phi_{K_0^*}^s(u)$ and $\phi_{K_0^*}^\sigma(u)$ are the twist-3 light-cone distribution amplitudes of the scalar meson ${K_0^*}$.  Further details of these can be found in the Appendix \ref{AppA}. Correspondingly the heavy meson projector used in this calculation are:
\begin{equation}
\mathcal{M}^{B}_{jk}
=
-\frac{i f_B m_B}{4}
\left[
\frac{1+\slashed{v}}{2}
\left\{
\phi_+^B(l_+)\,\slashed{n}_+
+
\phi_-^B(l_+)
\left(
\slashed{n}_-
-
l_+\,\gamma^\nu_\perp
\frac{\partial}{\partial l^\nu_\perp}
\right)
\right\}
\gamma_5
\right]_{jk}
\Bigg|_{\,l=\frac{l_+}{2}n_+}\;,
\label{B project}
\end{equation}
where, $\phi_+^B(l_+)$  and $\phi_-^B(l_+)$ are the distribution amplitudes of the $B$-meson which are discussed in Appendix \ref{AppB}.
The hard scattering amplitude in Feynman gauge takes the form 
\begin{equation}
\mathcal{T}_{ijkl}^\Gamma=\left[\Gamma\frac{m_b(1+\slashed{v})+\slashed{l}-\slashed{k_2}}{(m_bv+l-k_2)^2-m^2_b}\gamma_\mu+\gamma_\mu\frac{\slashed{k}_1+\slashed{k}_2-\slashed{l}}{(k_1+k_2-l)^2}\Gamma\right]_{i{}_j}\frac{1}{(l-k_2)^2}[\gamma^\mu]_{k{}_l}\;.
\label{hseq3}
\end{equation}
The gluon momenta give $(l-k_2)^2 \sim -2\,l_+\,\bar{u}\,E_F$ and the numerator of the first term gives $\slashed{l}-\slashed{k}_2 \sim -\bar{u}E_F\slashed{n}_-$, both these terms are of order $m_B\Lambda_{\rm QCD}$ and they contribute together with the leading term $m_b(1+\slashed{v})$. 
The total contribution from hard-gluon exchange by neglecting terms of order $\Lambda_{\rm QCD}/m_B$ from the hard-spectator kernel is given by
\begin{equation}
\mathcal{T}_{ijkl}^\Gamma\simeq\left[\Gamma\frac{m_b(1+\slashed{v})-\bar{u}E_F\slashed{n}_-}{4\bar{u}^2l_+m_bE^2_F}\gamma^\mu+\gamma^\mu\frac{E_F\slashed{n}_- -\slashed{l}}{4\bar{u}l^2_+E^2_F}\Gamma\right]_{i{}_j}[\gamma^\mu]_{k{}_l}\;.
\label{hseq4}
\end{equation}
The term $m_b(1+\slashed{v})$ exhibits a logarithmic divergence at the limit $\bar{u}\to 0$, since the DA $\phi(u)$ vanishes linearly at leading twist. These endpoint divergences do not break the symmetry relations and can be absorbed in the soft form factors in the factorization scheme in Eq.~(\ref{fact}). This can be verified from the current structure defined in Eq.~(\ref{AA}). 

In the present analysis we employ the twist-2 distribution amplitude which is the leading order term, whereas the twist-3 distribution amplitudes are suppressed by a factor of $1/m_B$. However, this suppression is compensated by a factor appearing in $\mathcal{M}^{K_0^*}$ in the endpoint limit $\bar{u}\to 0$. Consequently, these terms contribute at leading order to the soft form factors $\xi_{K_0^*}$. The details of the leading order terms appearing in the hard scattering kernel along with the underlying power counting framework can be found in \cite{Beneke:2000wa}.

The contributions arising from hard-spectator interactions are presented in detail for the case of $f_-(q^2)$. Similar calculations can be carried out for $f_T\left(q^2\right)$ as well. These contributions are denoted by $\Delta F_P$ and are expressed as 
\begin{equation}
\Delta F_P=\frac{8\pi^2f_Bf_{K_0^*}}{N_cm_B}\langle l_+^{-1}\rangle_+\langle\bar{u}^{-1}\rangle_P\;.
\end{equation}
Evaluation of these corrections requires the moments of the distribution amplitude $\langle l_+^{-1}\rangle_+$ and $\langle\bar{u}^{-1}\rangle$ for $B$-meson and the scalar meson $ K^*_0$, respectively, given as 
\begin{equation}
 \langle l_+^{-1}\rangle_+   =\int dl_+\frac{\phi^B_+(l_+)}{l_+}\;,
 \label{bbar}
\end{equation}
\begin{equation}
\langle\bar{u}^{-1}\rangle   =\int du\frac{\phi^{K_0^*}(u)}{\bar{u}}\;.
\label{ubar}
\end{equation}

With the help of these expressions, we can calculate the hard spectator correction of $B\to  K^*_0$ form factors $\Delta f_i$. However, the renormalization condition in Eq.(\ref{para}) implies $\Delta f_+=0$ by definition. The remaining two form factors are then given as
\begin{equation*}
   \Delta f_-=\frac{m_B^2-m_{K_0^*}^2}{2E_F m_B}\Delta F_P\;, \hspace{1cm}
  \Delta f_T=-\frac{m_B+m_{K_0^*}}{2E_F}\Delta F_P \;.
\end{equation*}
The total  $\mathcal{O}(\alpha_s)$ correction is the sum of the vertex and hard-spectator corrections. Combining these contributions will yield:
\begin{align}
f_{-}(q^2)
&= -\xi_{K_0^*}\Bigg[
\frac{\left(1 + \frac{m_{K_0^*}^2}{m_B^2}\right)}{\left(1 - \frac{2m_{K_0^*}^2}{m_B^2}\right)}
- \frac{\alpha_s C_F E_F}{4\pi}
\Bigg\{
\frac{m_{K_0^*}^4+4 \Delta E m_{K_0^*}^2}{2 E^2 \Delta}
\left(
\dfrac{1-\dfrac{m_B}{2 E_f} L}{2 q^2} -\dfrac{2-L^{\prime\prime}}{2 q^2}
\right) \nonumber \\
&\qquad
- \frac{m_{K_0^*}^2}{m_B E\,\Delta}
\left(
\frac{2m_B^2}{m_B^2 - q^2}L- L^\prime
- \ln\!\frac{m_b^2}{\mu^2} + 3
\right)
+ 4m_B \left(\frac{1 - L^\prime}{q^2}\right)
- \frac{m_{K_0^*}^4}{ E^2\Delta}
\left(
\frac{1 - \frac{m_B }{2E_F}L}{2 q^2}
\right)
\Bigg\}
\Bigg]
\nonumber\\[4pt]
&\quad
+ \frac{\alpha_s C_F}{4\pi}
\left(
\frac{m_B^2-m_{K_0^*}^2}{2E_F m_B}
\right)
\Delta F_P \;,
\label{symf}
\end{align}
and
\begin{align}
f_T(q^2)=&\xi_{K_0^*}\Bigg[
\left(\frac{m_B+m_{K_0^*}}{m_B}\right)
\left(\frac{m_B^2+m_{K_0^*}^2}{m_B^2-m_{K_0^*}^2}\right)
\nonumber\\[4pt]
&+\frac{\alpha_s C_F E_F}{4\pi}
\frac{ m_B(m_B+m_{K_0^*}) }{(E_F m_B - m_{K_0^*}^2)}
\Bigg\{
\left(
\frac{1}{m_B}+\frac{m_{K_0^*}^2}{2E\Delta m_B}
-\frac{m_{K_0^*}^2}{E m_B^2}
-\frac{m_{K_0^*}^4}{4E^2\Delta m_B^2}
\right)\frac{2m_B^2}{m_B^2-q^2}L
\nonumber\\[6pt]
&+\left(
-4\Delta-\frac{2m_{K_0^*}^2}{E}
+\frac{4m_{K_0^*}^2}{m_B}
\right)\frac{1-\frac{m_B}{2E_F}L}{2q^2}
+\left(
4m_B+\frac{m_B m_{K_0^*}^2}{E\Delta}
-\frac{3m_{K_0^*}^2}{E}
-\frac{m_{K_0^*}^4}{2E^2\Delta}
\right)\frac{1-L^\prime}{2q^2}
\nonumber\\[6pt]
&+\left(
-2\Delta+\frac{m_{K_0^*}^4}{2E\Delta m_B}
-\frac{m_{K_0^*}^4}{4E^2\Delta}
+\frac{m_{K_0^*}^2}{m_B}
-\frac{m_{K_0^*}^2}{E}
\right)\frac{2-L^{\prime\prime}}{2q^2}
\Bigg\}\Bigg]-\frac{\alpha_s C_F}{4\pi}\left(\frac{m_B+m_{K_0^*}}{2E_F}\right)\Delta F_P\;.
\label{symft}
\end{align}
\section{Numerical Analysis and Applications}\label{Tanalysis}
We now turn to the numerical analysis, focusing not only on the dependence of the form factors on momentum transfer $q^2$ given in Eq. (\ref{symf}) and Eq. (\ref{symft}), but also on their phenomenological implications. In particular, evaluating several physical observables including the decay rate and polarization asymmetries.
\subsection{Form Factors Analysis}
As discussed earlier, the correction in form factors receives contributions from the vertex diagrams having the uncertainties originating from the renormalization scale of $\alpha_s$. However, the hard-scattering corrections are subjected to large theoretical uncertainties in their computation. These uncertainties arise not only from the respective decay constants but also from the inverse moments defined in Eq. (\ref{bbar}) and Eq. (\ref{ubar}). It should be noted that the scale for the hard-scattering correction is $(m_B\Lambda_{QCD})^{1/2}$ and all quantities are evaluated at the scale of $\mu=1.47$ GeV \cite{Beneke:2000wa}. 

\textit{Meson decay constants}: Using the decay constants of $B$- and $K^*_0\left(1430\right)$ mesons as $f_B= 0.195 \pm 0.01 $ GeV \cite{Gelhausen:2013wia} and $f_{K^*_0}=0.427$ GeV \cite{Du:2004ki,Yin:2025}, respectively, the uncertainties in the evaluation of hard scattering corrections arising from these are estimated to be at the level of $\pm15\%$.

\textit{Light-cone distribution amplitudes}: The light-cone distribution amplitudes (LCDAs) play a central role in the hard-scattering amplitudes. In particular, the inverse moment satisfies 
$\langle l_+^{-1} \rangle_+ \sim \mathcal{O}(1/\Lambda_{\mathrm{QCD}})$; in the present analysis, we adopt the numerical value of moment of distribution amplitude of $B$-meson as $\langle l_+^{-1} \rangle_+ = (0.35\,\mathrm{GeV})^{-1}$. 
For the $K_0^*(1430)$ meson, the LCDA is expanded in terms of Gegenbauer moments, the numerical value of moment of distribution amplitude of $K_0^*(1430)$ is $\langle\bar{u}^{-1}\rangle$=(1.422) the details of which are provided in Appendix~\ref{AppA}.

\textit{Soft form factor}: To determine the soft form factor $\xi_{K_0^*}(E_F)$ in the large-recoil region required for evaluating the form factors $f_+$, $f_-$, and $f_T$ - we employ results from light-cone sum rules (LCSR) for $B \to K_0^* \ell^+ \ell^-$ decays~\cite{Wang:2008da,Aslam:2010mc,Khosravi:2024zaj}. 
We note that these studies do not include perturbative corrections. 
The value of $\xi_{K_0^*}$ is extracted from the form factor $f_+(q^2=0)$ using Eq.~(\ref{para}), yielding $\xi_{K_0^*}(0) = 0.903$.

\begin{figure}[H]
    \centering
\includegraphics[width=8cm, height=6cm]{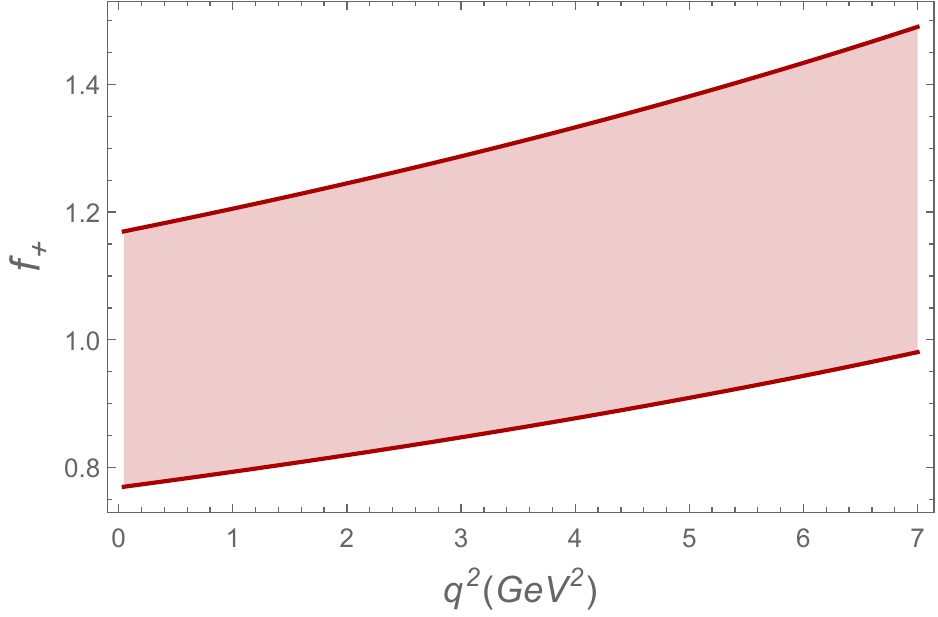}
\includegraphics[width=8cm, height=6cm]{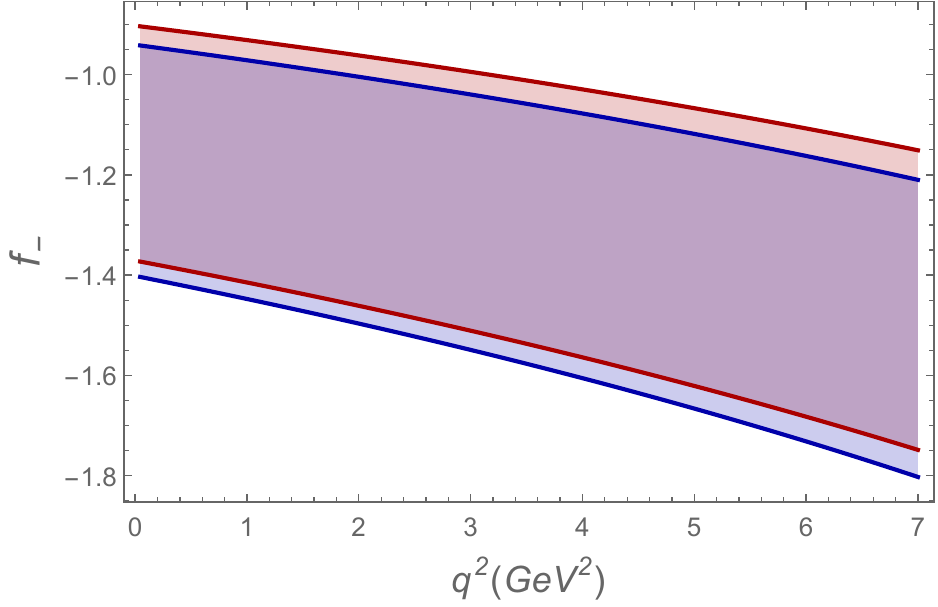}
\includegraphics[width=8cm, height=6cm]{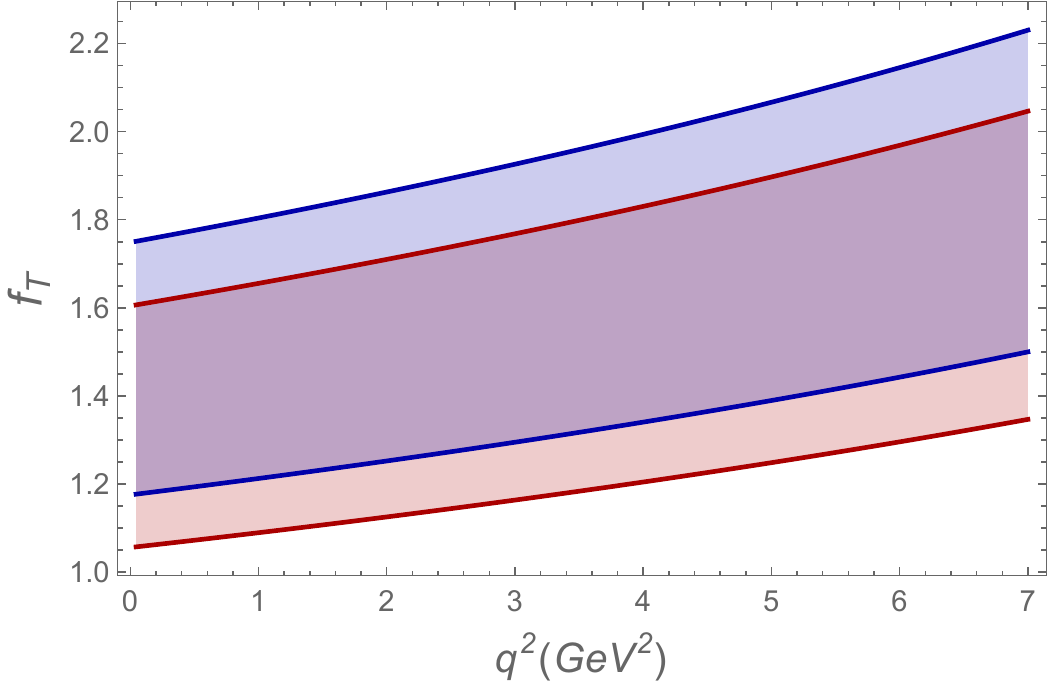}
\caption{Form factors are plotted as function of momentum transfer $q^2$. The  uncertainties are calculated at $q^2=0$ and the extrapolated with $q^2$. The red band represents the trend without symmetry breaking effects, and the blue band is obtained when the symmetry breaking corrections are incorporated.}
\label{fig:plotFminus&t}
\end{figure}

The $q^2$-dependence of $\xi_{K_0^*}$ is modeled using single- and double-pole parameterizations, following Ref.~\cite{Aslam:2010mc}, over the kinematic range $0 < q^2 < (m_B - m_{K_0^*})^2$. 
The non-perturbative parameters $a_i$ and $b_i$ are determined within the LCSR framework in the low-$q^2$ region. The corresponding errors associated with them, together with the uncertainties in the soft form factor at $q^2=0$ and other input parameters, are propagated to obtain the uncertainty bands for the form factors and related physical observables. Since the large-recoil symmetry relations are valid only for $q^2 \ll m_B^2$, the analysis is restricted to $q^2 \lesssim 7\,\mathrm{GeV}^2$.
Using $\alpha_s = 0.34$ and the renormalization scale $\mu = 1.47\,\mathrm{GeV}$, the resulting form factors are computed and shown as functions of $q^2$ in Fig.~\ref{fig:plotFminus&t}.re
Using the uncertainties reported in \cite{Wang:2008da,Aslam:2010mc,Khosravi:2024zaj} along with other input parameters discussed above, we obtain bands as a function $q^2$. In these plots the red band represents the form factors without symmetry corrections, while the blue band corresponds to the same form factors after including the symmetry breaking corrections. As can be observed from the plot $f_{-}(q^2)$ exhibits the pronounced effects across the entire range of $q^2$ under consideration. For $f_T(q^2)$, normalized at $\mu=m_b$, it can be seen that in most of the $q^2$ region the two bands overlap significantly showing that the symmetry breaking corrections are largely obscured by the various uncertainties. It is to emphasize here again that major uncertainty lies in hard spectator corrections due to LCDA of the $B$-meson. In past, due to the non-availability of constraints on inverse moments this
uncertainty could rise as high as $\pm 50\%$ \cite{Beneke:2000wa}. These uncertainties were constrained by BABAR analysis of $B \rightarrow \gamma \ell \nu_\ell$ \cite{BaBar:2009pvj} at small recoil and can further be improved by a similar analysis by BABAR for large recoil radiative decay. Their analysis was further improved in \cite{Beneke:2011nf} as the former does not consider highly energetic photons and radiative/power corrections. In
context of \cite{Beneke:2011nf}, we expect more uncertainty at large recoil than at the small recoil. 

\subsection{Applications}
\begin{figure}[H]
    \centering
    \includegraphics[width=10cm, height=6cm]{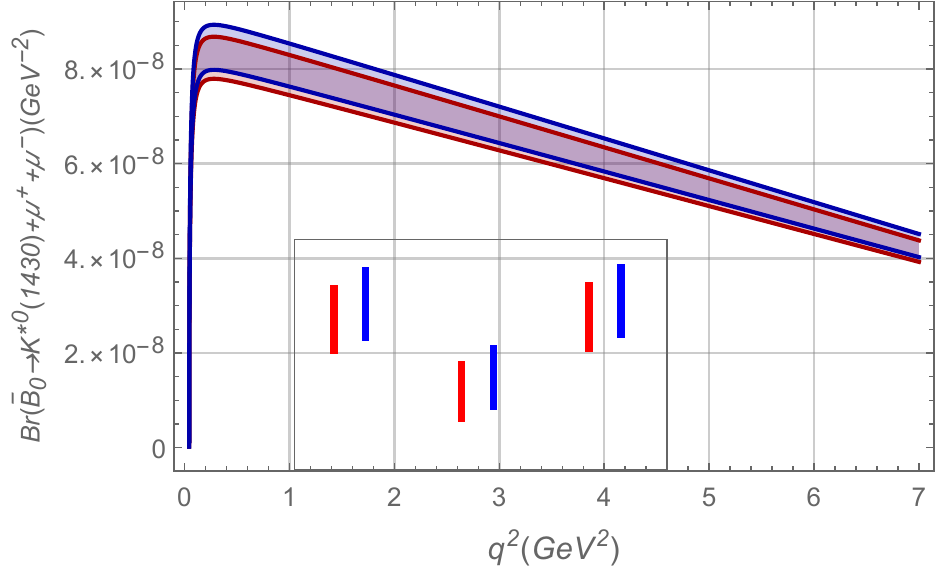}
\caption{The branching ratio of $B\to K^*_0\mu^+\mu^-$ with respect to $q^2$.}
    \label{fig:branching}
\end{figure}

It is always valuable to investigate the impact of different corrections to form factors derived above by examining relevant observables. In this context, decay amplitude and polarization asymmetries provide interesting examples. In this section, the effect of symmetry corrections to the form factor will be analyzed in $B\to K_0^*\left(1430\right)\ell^+\ell^-$ decay in the SM. The decay of the $B$-meson to a light scalar meson is induced by the flavor changing neutral current transition $b\to s\ell^+\ell^-$. Within SM, it is described by effective Hamiltonian given in Eq. (\ref{H11}) indicating the contributions that factorizes into hadronic matrix elements. Since in general, not all contributions can be fully expressed in terms of form factors alone, as non-factorizable hard-scattering terms also arise  (see e.g., \cite{Beneke:2001at}). However, in the present framework, these non-factorizable effects are not included. Thus, the decay amplitude can be summarized as 
\begin{equation}
   \mathcal{M}_{B\to K_0^*\ell^+\ell^-}=-\frac{G_F\alpha}{2\sqrt{2}\pi}V_{tb}V_{ts}^*[T_{\mu}^1(\bar{\ell}\gamma^\mu \ell)+T_{\mu}^2(\bar{\ell}\gamma^\mu\gamma_5 \ell)] 
\end{equation}
where $T_\mu^1$ and $T_\mu^2$ are the hadronic parts of the decay amplitude defined as 
\begin{equation}
 T_\mu^1=-C_9^{\text{eff}}f_+(q^2)p_\mu+\frac{4m_b}{m_B+m_{K^*_0}}C_7^{\text{eff}}f_T(q^2)p_\mu
\end{equation}
\begin{equation}
  T_\mu^2=-C_{10} (f_+(q^2)p_\mu+f_-(q^2)q_\mu), 
\end{equation}
\textbf{The differential decay rate}:\\
The differential decay rate of this transition is given as \cite{ParticleDataGroup:2008zun}:
\begin{align}
  \dfrac{d\Gamma}{dq^2}&\propto \Bigg[ \Big |-C_9^{\text{eff}}f_+(q^2)p_\mu+\frac{4m_b}{m_B+m_{K^*_0}}C_7^{\text{eff}}f_T(q^2)p_\mu \Big|^2\left(2m_l^2+q^2\right)\lambda + 12m_l^2q^4\Big|-C_{10}f_-(q^2)q_\mu\Big|^2\nonumber \\ 
 &+ \Big|-C_{10} f_+(q^2)p_\mu\Big|^2\left\{(2m_l^2+q^2)(m_B^4-2m_B^2m^2_{K_0^*}-2q^2m^2_{K_0^*})+(m^2_{K_0^*}-q^2)^2+2m_l^2(m^4_{K_0^*}+10m^2_{K_0^*}+q^4)\right\}\nonumber \\
  &+12q^2m_l^2(m_B^2-m^2_{K_0^*}-q^2)\left|C_{10}\right|^2 p\cdot q\left(f_-(q^2)f_+^*\left(q^2\right)+f_-^*(q^2)f_+\left(q^2\right)\right) 
  \Bigg]\;,
\end{align}
where
\begin{equation}
\lambda=\lambda(m^2_B,m^2_{K_0^*},q^2)=m_B^4+m^4_{K_0^*}+q^4-2m_B^2m^2_{K_0^*}-2m^2_{K_0^*}q^2-2q^2m_B^2\;.
\end{equation}
The branching ratio is given as a function of momentum transfer $q^2$ in Fig.(\ref{fig:branching}). In the plots, the red curves represents the branching ratio at tree level and the blue curves corresponds to the branching ratio after including the symmetry breaking corrections. The band represents the uncertainties already discussed in the previous section. As can be observed that the inclusion of symmetry breaking corrections in form factors leads to only a small shift compared to the case without these corrections. In the inset bar plots illustrates the effect of form factors on observables across three different $q^2$ bins $(q^2_{\text{min}}-2)$, $(2-4)$ and $(4-7)$ $\text{GeV}^2 $.
\subsubsection{Lepton polarizations}

\begin{figure}[H]
    \centering
    \includegraphics[width=8.5cm, height=5cm]{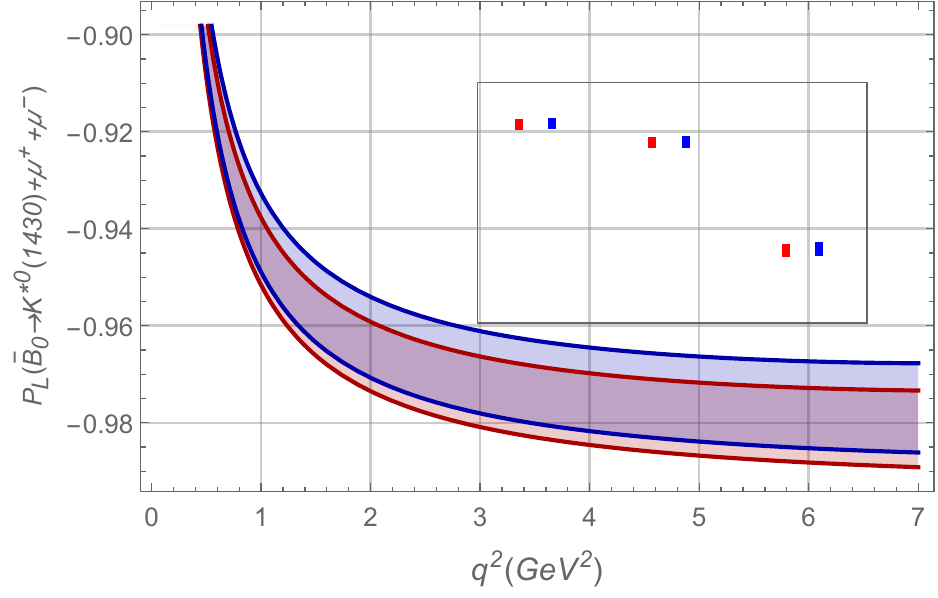}
    \includegraphics[width=8.5cm, height=5cm]{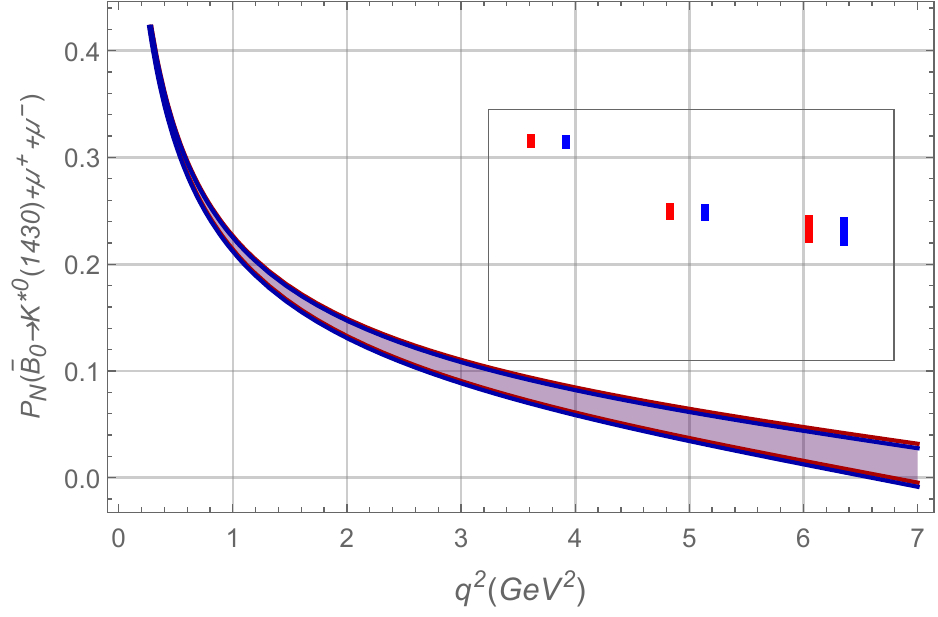}
    \caption{The lepton polarization of $B\to K^*_0\mu^+\mu^-$ with respect to $q^2$.}
    \label{fig:placeholder}
\end{figure}

The lepton pair produced in this decay has longitudinal, normal and transverse components of polarization. The expression for longitudinal polarization can be summarized as:
\begin{align}
    P_L(q^2)& \propto \left(1-\frac{4m_l^2}{q^2}\right)\Bigg[(-C_9^{\text{eff}}f_+(q^2)p_\mu+\frac{4m_b}{m_B+m_{K^*_0}}C_7^{\text{eff}}f_T(q^2)p_\mu))(-C_{10} f_+(q^2)p^\mu)^* \nonumber \\ 
    &+(-C_9^{\text{eff}}f_+(q^2)p_\mu+\frac{4m_b}{m_B+m_{K^*_0}}C_7^{\text{eff}}f_T(q^2)p_\mu)^*(-C_{10} f_+(q^2)p^\mu)\Bigg]
\end{align}
The normal lepton polarization of the $B\to K^*_0\ell^+\ell^-$ is calculated through the following expression: 
\begin{equation}
\begin{aligned}
P_N(q^2) \propto \left(1 - \frac{4 m_l^2}{q^2}\right)
\Bigg[
&\left(
- C_9^{\mathrm{eff}} f_+(q^2)p_\mu
+ \frac{4 m_b}{m_B + m_{K_0^*}} C_7^{\mathrm{eff}} f_T(q^2)p_\mu
\right)
\left(- C_{10} f_+(q^2)p^\mu\right)^*
\\
&+
\left(
- C_9^{\mathrm{eff}} f_+(q^2)p_\mu
+ \frac{4 m_b}{m_B + m_{K_0^*}} C_7^{\mathrm{eff}} f_T(q^2)p_\mu
\right)^*
\left(- C_{10} f_+(q^2)p^\mu\right)
\\
&- 2 q^2
\left(
- C_9^{\mathrm{eff}} f_+(q^2)p_\mu
+ \frac{4 m_b}{m_B + m_{K_0^*}} C_7^{\mathrm{eff}} f_T(q^2)p_\mu
\right)^*
\left(- C_{10} f_-(q^2) q^\mu\right)
\\
&+
\left(
- C_9^{\mathrm{eff}} f_+(q^2)p_\mu
+ \frac{4 m_b}{m_B + m_{K_0^*}} C_7^{\mathrm{eff}} f_T(q^2)p_\mu
\right)
\left(- C_{10} f_-(q^2) q^\mu\right)^*
\Bigg]
\end{aligned}
\end{equation}
Fig.(\ref{fig:placeholder}) illustrate the behavior of $P_L$ and $P_N$, where color coding is same as used for the branching ratio. The plots are overlapping in the low $q^2$ region while a significant deviation is observed in the behavior of the lepton polarization at intermediate $(q^2)$ region.  
\section{Conclusion}\label{conclusion}
In this work, we have computed the symmetry-breaking corrections to the form factors governing the rare decay $B \to K_0^* \ell^+ \ell^-$ at one-loop order. These corrections become particularly relevant in the large-recoil region, $q^2 \sim 1$--$7\,\mathrm{GeV}^2$. The structure of these effects is described within the QCD factorization framework~\cite{Beneke:2003pa}, where short-distance contributions from heavy degrees of freedom and hard virtual corrections are treated perturbatively, while long-distance dynamics associated with light quarks and gluons are encoded in nonperturbative hadronic matrix elements parameterized by form factors.

In the heavy-quark and large-recoil limit, the three independent form factors $f_+(q^2)$, $f_-(q^2)$, and $f_T(q^2)$ relevant for $B \to K_0^* \ell^+ \ell^-$ transitions reduce to a single universal soft form factor $\xi_{K_0^*}$. Corrections to these symmetry relations arise from hard-gluon interactions and can be systematically classified into vertex and hard-spectator contributions. The vertex corrections are computed at $\mathcal{O}(\alpha_s)$ through matching of the effective theory onto full QCD, while the hard-spectator contributions are evaluated using the light-cone distribution amplitudes of the participating mesons.

Our numerical analysis shows that the inclusion of symmetry-breaking corrections leads to deviations of about $4\%$ in $f_-(q^2)$ and $10\%$ in $f_T(q^2)$ relative to their symmetry-limit values. These modified form-factor relations at large recoil have a noticeable impact on phenomenological observables. In particular, we find that the branching ratio receives a correction of approximately $3\%$, while the normal lepton polarization asymmetry $P_N$ is also affected at the level of $\sim 3\%$. The longitudinal polarization $P_L$ remains largely unchanged, whereas the transverse polarization $P_T$ does not receive any significant contribution from symmetry-breaking effects in the present framework. It should be noted here that the uncertainties from the meson decay constants are $\pm 15\%$. The parameter enters through sub-leading contributions and convolution integrals in the hard-scattering kernel. Thus, the dominant contributions are governed by form-factor terms, while the dependence on $f_B$ and $\langle l_+^{-1} \rangle_+$ is either power-suppressed or enters in combination with other hadronic quantities inside convolution integrals. Consequently, despite the relatively large uncertainties in the input parameters, their net impact on the final observable remains at the level of a few percent. Therefore, due to the relatively small magnitude of these symmetry-breaking corrections, any sizable deviations observed from the SM predictions in these decays would be a clear hint of potential New Physics effects. 

In future lattice QCD calculations are expected to play a crucial role in reducing the uncertainties arising from hadronic inputs. Thus, more precise lattice determinations of $f_B$ and improved constraints on the low moments of the B-meson distribution amplitude , either directly or via related HQET matrix elements, would significantly reduce the parametric uncertainty in the hard-scattering contributions. This would allow a more precise isolation of subleading power corrections in $B \rightarrow K^*_0(1430)$ transitions. On the experimental side, advancement in the precise measurements of different observables and LFU-sensitive ratios will further tighten the constraints on form factors, which will help disentangle hadronic effects from possible NP contributions. Specifically, improved data on branching ratios and angular observables will reduce the allowed range of input parameters. As a result, future progress both theoretically and experimentally will substantially reduce these uncertainties and enhance the sensitivity to such subleading effects.

\appendix
\section{Light cone Distribution Amplitude of $K^*_0$(1430)}
\label{AppA}
The light cone distribution amplitudes of the scalar meson at the twist-2 $\phi_S(u)$ and the twist-3 $\phi^s_S(u)$ and  $\phi^\sigma_S(u)$ are given as
\begin{equation}
\begin{aligned}
\langle S(p)|\,\bar{q}_2(z_2)\gamma_\mu q_1(z_1)\,|0\rangle
&= f_S p_\mu \int_0^1 du\;
e^{i(up\cdot z_2 + \bar{u}p\cdot z_1)}\,
\phi_S(u),
\\[6pt]
\langle S(p)|\,\bar{q}_2(z_2) q_1(z_1)\,|0\rangle
&=\bar f_S m_S \int_0^1 du\;
e^{i(up\cdot z_2 + \bar{u}p\cdot z_1)}\,
\phi_S^s(u),
\\[6pt]
\langle S(p)|\,\bar{q}_2(z_2)\sigma_{\mu\nu} q_1(z_1)\,|0\rangle
&= -\bar f_Sm_S (p_\mu z_\nu - p_\nu z_\mu)
\int_0^1 du\;
e^{i(up\cdot z_2 + \bar{u}p\cdot z_1)}\,
\frac{\phi_S^\sigma(u)}{6}.
\end{aligned}
\label{scalar_lcda}
\end{equation}
As a single matrix element, above LCDAs of Eq. (\ref{scalar_lcda}) can be combined as
\begin{equation}
\langle S(p)|\,\bar{q}_{2\beta}(z_2)\,q_{1\alpha}(z_1)\,|0\rangle
=
\frac{\bar f_S}{4}
\int_0^1 du\;
e^{i(up\cdot z_2 + \bar{u}p\cdot z_1)}
\left\{
\not{p}\,\phi_S(u)
+
m_S
\left(
\phi_S^s(u)
-
\sigma_{\mu\nu} p^\mu z^\nu
\frac{\phi_S^\sigma(u)}{6}
\right)
\right\}_{\alpha\beta}
\end{equation}
The general expression for twist-2 LCDA for the scalar mesons is given by
\begin{equation}
    \phi_S(u,\mu)=6u\bar{u}\left[B_0(\mu)+\sum^\infty_{n=1} B_n(\mu)C^{3/2}_n\left( 2u-1 \right)\right]
    \label{app1}
\end{equation}
where $\bar{u}=1-u$ and $S={K^*_0}$ \cite{Han:2013zg,Hai:2006}. The normalization conditions 
gives $ B_0=\mu_S^{-1}$. The coefficients $B_n(\mu)$ are the Gegenbauer moments, while $C^{3/2}(2u-1)$ are the Gegenbauer polynomials, both depending on the renormalization scale $\mu$. Due to the asymmetric nature of scalar meson LCDA, only odd Gegenbauer moments contribute, leading to the vanishing of even moments, i.e., $B_2 = B_4 = 0$. Details of these can be found in \cite{Wang:2008da}. The first few Gegenbauer polynomials are given by
\begin{equation}
C_1^{3/2}(2u-1)=3(2u-1), \qquad
C_2^{3/2}(2u-1)=\frac{3}{2}\left[5(2u-1)^2-1\right], \qquad \cdots
\end{equation}
Neglecting the higher-order contributions and retaining only the leading nontrivial term, Eq. (\ref{app1}) reduces to
\begin{equation}
     \phi_{K^*_0}(u,\mu)=6u\bar{u}[B_0(\mu)+ B_1(\mu) C_1^{3/2}(2u-1)]\;.
\end{equation}
This is the final truncayed form of the LCDA for the $K^*_0$ meson. The leading-twist moment for the scalar meson can then be obtained by integrating the above expression, as referenced in Eq.~(\ref{ubar}).
\section{$B$-meson momentum space projector}
\label{AppB}
In this section, we evaluate the projection operator of the $B$-meson as defined in Eq.~(\ref{B project}). The derivation begins with the two-particle light-cone matrix element in coordinate space. To proceed, we introduce the distribution amplitudes $\phi_\pm^B(t)$ through the Lorentz decomposition of the corresponding light-cone matrix element, which can be written as follows:
\begin{equation}
    \langle0|\bar{q}_\beta(z)P(z,0)b_\alpha(0)|\bar{B}(p\rangle)=-\frac{if_Bm_B}{4}\Bigg[ \frac{1+\not{v}}{2}\{2\phi_+^B(t)+\frac{\phi_-^B(t)-\phi_+^B(t)}{t}\not{z}\}\gamma_5\Bigg]_{\alpha\beta}
    \label{Bproj}.
\end{equation}
We now impose the light-cone condition $z^2 = 0$, under which the variable $t$ is defined as $t = v \cdot z$. The momentum of the $B$-meson is expressed as $p = m_B v$, where $m_B$ denotes the mass of the $B$-meson. The path-ordered exponential can then be written as 
\begin{equation}
    P(z_2,z_1)=P\text{exp}\Bigg(ig_s\int^{z_1}_{z_2}dz^\mu A_\mu(z)\Bigg).
\end{equation}
The momentum space projector in Eq. (\ref{Bproj}) is most compatible with the Lorentz invariance and in the heavy quark limit. The first factor is chosen in such a way that for $z=0$ it gives
\begin{equation}
    \langle0|\bar{q}_\beta[\gamma^\mu\gamma_5]_{\beta\alpha} b_\alpha|\bar{B}(p\rangle)=if_Bm_Bv^\mu,
\end{equation}
if $\phi_+^B(t=0)=\phi_-^B(t=0)=1$. $\phi_+^B$ is the leading twist and $\phi_-^B$ is the sub leading twist term. Now in Eq. (\ref{Bproj}) for $M(z)$ to be the matrix element and $A(z)(A(l))$ is the hard scattering amplitude in momentum space, Eq. (\ref{B project}) can be achieved by using the identity 
\begin{equation}
    \int d^4zM(z)A(z)=\int\frac{d^4l}{(2\pi)^4}A(l)\int d^4ze^{-ilz}M(z)\equiv\int^\infty_0dl_+M^BA(l)\Bigg|_{l=\frac{l_+}{2}n_+},
    \label{Ahard}
\end{equation}
with the components of $l^\mu$ are decomposed as 
\begin{equation}
    l^\mu=\frac{l_+}{2}n^\mu_++\frac{l_-}{2}n^\mu_-+l^\mu_\perp
\end{equation}
and the coordinate functions $\phi_\pm^B(t)$ in momentum space are written as 
\begin{equation}
    \phi_\pm^B(t)\equiv\int^\infty_0dwe^{-iwt}\phi_\pm^B(w).
    \label{phiB}
\end{equation}
The hard scattering amplitude $A(l)$ of the light meson moving in the $n_-$ direction in the heavy quark limit does not depends on $l_-$. So 
\begin{equation}
    A(l)=A^0(l_+)+l^\mu_\perp A^1_\mu(l_+)+O(1/m_B).
\end{equation}
and the derivative after ignoring the $l_-$ term is 
\begin{equation}
    \frac{\partial}{\partial l_\mu}=n_-^\mu\frac{\partial}{\partial l_\mu}+\frac{\partial}{\partial l_{\perp\mu}}.
    \label{lmu}
\end{equation}
Combining Eq. (\ref{phiB}) and Eq. (\ref{lmu}) in Eq. (\ref{Ahard}) one gets a momentum space projector which is a two particle light cone projector involving only a $b$-quark and a light spectator quark and holds when the three-particle contribution is neglected, it is written as
\begin{equation}
    M^B_{\beta\alpha}=-\frac{if_Bm_B}{4}\left[
\frac{1+\not{v}}{2}
\left\{
\phi_+^B(w)\,\not{n}_+
+
\phi_-^B(w)
\left(
\not{n}_-
-
l_+\,\gamma^\nu_\perp
\frac{\partial}{\partial l^\nu_\perp}
\right)
\right\}
\gamma_5
\right]_{\beta\alpha}
\Bigg|_{\,l=\frac{l_+}{2}n_+}.
\end{equation}

\end{document}